\documentstyle[prb,preprint,eqsecnum,aps]{revtex}
\begin{document}
\draft
\title{Structure Factor and Electronic Structure \\
of Compressed Liquid Rubidium
}
\author{Junzo Chihara}
\address{Advanced Photon Research Center, Japan Atomic Energy Research Institute\\
Tokai, Ibaraki 319-11, Japan}
\author{Gerhard Kahl}
\address{Institut f\"ur Theoretische Physik and CMS, 
Technische Universit\"at Wien\\
Wiedner Hauptstra\ss e 8-10, A-1040 Wien, Austria}
\date{\today}
\renewcommand{\baselinestretch}{2.0}
\maketitle
\begin{abstract}
We have applied the quantal hypernetted-chain equations in combination 
with the Rosenfeld bridge-functional to calculate the atomic and the
electronic structure of compressed liquid-rubidium under high pressure
(0.2, 2.5, 3.9, and 6.1 GPa);  
the calculated structure factors are in good agreement with experimental 
results measured by Tsuji {\it et al.} along the melting curve.
We found that the Rb-pseudoatom remains under these high pressures almost 
unchanged with respect 
to the pseudoatom at room pressure; thus, the effective ion-ion 
interaction is practically the same for all pressure-values. We observe
that all structure factors calculated for this pressure-variation 
coincide almost into a single curve if wavenumbers are
scaled in units of the Wigner-Seitz radius $a$ although no
corresponding scaling feature is observed in the 
effective ion-ion interaction. 
This scaling property of the 
structure factors signifies that the compression in liquid-rubidium is 
uniform with increasing pressure; in absolute $Q$-values this means 
that the first peak-position 
($Q_1$) of the structure factor increases proportionally to $V^{-1/3}$ 
($V$ being the specific volume per ion), as was experimentally
observed by Tsuji {\it et al}.
This scaling property comes from a specific feature 
characteristic for 
effective ion-ion potential of liquid rubidium: 
even if the effective liquid-rubidium potential 
is invariant under this pressure-variation with respect to 
the potential 
at room pressure, we can nevertheless observe this scaling property for 
the structure factors. 
This property is obviously characteristic for the potential of 
alkali metals and we have examined and confirmed this feature 
for the case of a liquid-lithium potential: starting from the 
liquid-lithium potential at room pressure we can easily find two
sets of densities and temperatures 
for which the structure factors become practically identical, when 
scaling $Q$ in units of $a$.
\end{abstract}

\section{INTRODUCTION}

Based on density-functional (DF) theory, 
we have derived in previous work
a set of integral equations, which allows to calculate
the ion-ion and the electron-ion correlations in a liquid metal 
or a plasma, consistent with the bound-electron structure of the ion using 
only the atomic number $Z_{\rm A}$ as input; these integral equations 
are named the quantal hypernetted-chain (QHNC) equations,\cite{hyd,QHNC} 
which are derived from the exact expressions for the ion-ion and 
electron-ion radial distribution functions (RDF) in an electron-ion mixture.
Up to now, we have applied this approach to 
liquid metallic hydrogen,\cite{hyd} lithium,\cite{QHLi} 
sodium,\cite{QHNa} 
potassium\,\cite{QHK} and aluminum\,\cite{QHAl} obtaining 
ion-ion structure factors in excellent agreement with experiments. 
Since a liquid metal can be considered as a very special type 
of a high-density plasma, we can expect
from the successful application of the QHNC-method to liquid metals,
that this approach is able to provide accurate results for a plasma.
In such a system, both
the ionic valency $Z_{\rm I}$ and the electron-ion interaction 
$v_{\rm eI}(r)$ may vary over a wide range as 
temperature and density are changed. Our method is in particular suited
to treat a plasma, since it is able to calculate the ionization $Z_{\rm I}$ and the electron-ion interaction 
in a self-consistent manner using the atomic number 
of the system as the only input data.

Recently, we have extended this formalism and 
have performed a first-principles molecular dynamics (MD) simulation
based on
the QHNC theory for alkali metals near the triple point: in this study 
those small deviations which were still observed between experimental results 
and QHNC data for the structure factor disappeared 
completely.\cite{fpMD}
Nevertheless, the calculation of the ion-ion RDF in an MD
simulation is rather time-consuming. 
Recently, Rosenfeld\cite{newbr} has proposed 
a new bridge-functional for hard-spheres; its construction requires only
fundamental measures of the hard-spheres. In combination with the
MHNC-approach -- by replacing the bridge-{\it function} by a bridge-{\it
functional} of the reference system -- it was found out 
that this method is able to give very accurate results for the structure
and thermodynamics of a large variety of one-component and binary liquid
systems.\cite{kahl}
Therefore, with replacement of the MD simulations by the MHNC method 
we can obtain accurate results for the ion-ion and electron-ion RDF's for 
dense fluids via the QHNC method 
using the Rosenfeld bridge-functional.

In a plasma, density and temperature vary over a wide range. Therefore, 
in order to check the applicability of the approach presented here 
to a strongly coupled plasma 
(where no reliable experimental data for particle-correlations are
available), 
it is important to examine to what extent the QHNC equations can describe 
a liquid metal 
in a wide range of densities and temperatures.
Recently, Tsuji {\it et al.} \cite{tsuji} measured the 
structure factors of liquid Rb 
at high pressures: 0.2, 2.5, 3.9 and 6.1 GPa. 
These pressure-values bring along a compression of liquid Rb: the
corresponding density values are estimated to be 
1.07, 1.41, 1.56 and 1.95 times the normal density, respectively.\cite{Morim}
Tsuji {\it et al.} observed that liquid Rb is uniformly compressed 
with increasing pressure. 
In this context it should be mentioned that Shimojo 
{\it et al.} \cite{Shimojo} performed first-principles molecular-dynamics 
simulation for liquid Rb under pressure (0, 2.5, and 6.1 GPa) and 
compared their results with the RDF's 
extracted from the structure factors measured by Tsuji {\it et al.}: 
they found that 
liquid Rb is compressed uniformly at 2.5 GPa, but that some  
deviation from uniform compression is observed at 6.1 GPa.
In the present work, using the QHNC method in combination with the
Rosenfeld bridge-functional for the reference hard-sphere system we have
calculated the structure factors of compressed 
rubidium, the pressure ranging from 0 to 6.1 GPa and compared these data
with the experimental results. 
We find excellent agreement with
these experimental data, although in two cases the density has to be
slightly readjusted. We confirm with our results the uniform compression
model. We find out that the effective ion-ion potentials are insensitive
to this pressure-variation, a feature which is obviously typical for
liquid alkali metals (we also confirm these observations for the case of
liquid-lithium). We finally observe that the structure factors coincide in one
single curve if $Q$ is scaled in units of the Wigner-Seitz radius $a$.

The paper is organized as follows: in the subsequent section we
briefly outline the QHNC method and give a few details about the
Rosenfeld bridge-functional. In Section III we discuss our results and
compare them with experimental data. The paper is concluded by a
summary.

\section{QHNC THEORY AND THE BRIDGE FUNCTIONAL}
In the present section, we give a brief outline of the QHNC theory 
and the integration of the Rosenfeld bridge-functional in an 
integral equation approach for a one-component fluid.

Let us consider a liquid metal or a plasma as a mixture of electrons and
ions interacting through pair potentials $v_{ij}(r)$ [$i,j=\mbox{e
or I\,}$]. In this mixture, the ion-ion and electron-ion RDF's $g_{i\rm I}(r)$ 
are identical with the ion- and electron-density distributions 
under the external potential caused by a fixed ion at the origin 
respectively, since the 
ions behave as a classical fluid in a liquid metal.\cite{QHNC} 
In general, DF theory enables us to provide exact expressions for the 
density distributions in an inhomogeneous system caused by an 
external potential. 
Hence, by applying DF theory to this mixture with densities $n_0^i$, 
we can derive 
exact expressions for the ion-ion and electron-ion RDF's 
in terms of direct correlation functions (DCF's) $C_{ij}(r)$ and 
bridge-functions $B_{i\rm I}(r)$ as follows:\,\cite{QHNC}
\begin{eqnarray}
g_{\rm II}(r) & = & \exp[-\beta U_{\rm I}^{\rm eff}(r)]\;,\\
g_{\rm eI}(r) & = & n_{\rm e}^{0f}(r|U_{\rm e}^{\rm eff})/n_0^e\;,
\end{eqnarray}
with
\begin{eqnarray}
U_i^{\rm eff}(r)  &\equiv&  v_{i\rm I}(r)-{1\over\beta} \left[ \sum_l
\int C_{il}(|{\bf r}-{\bf r}'|)n_0^l[g_{l\rm I}(r)-1]d{\bf r}'
+B_{i\rm I}(r) \right] \\
& = & v_{i\rm I}(r)-\Gamma_{i\rm I}(r)/\beta \;.
\end{eqnarray}
The wave equation for an electron under the
external potential $U_{\rm e}^{\rm eff}(r)$ is solved to provide 
the total electron-density distribution $n_{\rm e}(r)$ around a nucleus, 
which is divided into the bound-electron and free-electron parts: 
$n_{\rm e}(r|U_{\rm e}^{\rm eff})\equiv n_{\rm e}^{0b}(r|U_{\rm e}^{\rm eff})+n_{\rm e}^{0f}(r|U_{\rm e}^{\rm eff})$ by the criterion whether their eigenfunctions belong to bound- or free-levels.   
In Eq. (2.2), the free-electron part of the density distribution, 
$n_{\rm e}^{0f}(r|U_{\rm e}^{\rm eff})/n_0^e$, becomes identical to 
the electron-ion RDF; 
the bound-electron part $n_{\rm e}^{0b}(r|U_{\rm e}^{\rm eff})$ is taken to
form an ion and contributes to generate the electron-ion interaction 
$v_{\rm eI}(r)$.  These expressions 
for $g_{i\rm I}(r)$ can be rewritten in the form of a set of integral 
equations for a one-component fluid.\cite{QHNC} One of them is a usual integral equation for the DCF 
$C(r)$ of a one-component fluid:
\begin{equation}\label{eq:QHNCii}
C(r)=\exp[-\beta v_{\rm eff}(r)
  +{\it\gamma}(r)+B_{\rm II}(r)]-1-{\it\gamma}(r)\;,
\end{equation}
with an interaction $v_{\rm eff}(r)$ defined by 
\begin{equation}\label{eq:veffii}
\beta v_{\rm eff}(Q)\equiv\beta v_{\rm II}(Q)
 -{|C_{\rm eI}(Q)|^2 n_0^{\rm e} \chi_Q^0\over
 1-n_0^{\rm e} C_{\rm ee}(Q) \chi_Q^0 }\;;
\end{equation}
the other is an equation for the effective interaction 
$v_{\rm eff}(r)$, that is expressed in the form of an integral
equation for the electron-ion DCF $C_{\rm eI}(r)$:
\begin{equation}\label{eq:QHNCei}
\hat B C_{\rm eI}(r)
 =n_{\rm e}^{0f}\left(r|v_{\rm eI}
     -{\it\Gamma}_{\rm eI}/\beta-B_{\rm eI}/\beta\right)/n_0^{\rm e}
     -1-\hat B {\it\Gamma}_{\rm eI}(r)\;. 
\end{equation}
In these equations, $\chi_Q^0$ is the density response function of the
noninteracting electrons with an electron density $n_0^{\rm e}$  
and
${\it\gamma}(r)\equiv \int C(|{\bf r}-{\bf r}'|)n_0^{\rm I}[g_{\rm II}
(r')-1]d{\bf r}'$.
Furthermore, $\widehat B$ denotes an operator defined by
\begin{equation}
{\cal F}_{\!\bf Q}[\widehat B f(r)] \equiv \chi_Q^0\!\int e^{i{\bf Q}\cdot{\bf r} }f(r)d{\bf r} \;.
\end{equation} 
It should be kept in mind that the electron-ion DCF in 
Eq.~(\ref{eq:veffii}) plays the role of a nonlinear pseudopotential, which 
takes into account nonlinear electron-accumulations around an ion 
and the influence of other ions in the form of a linear-response expression; 
if the electron-ion DCF $C_{\rm eI}(r)$ is replaced by a usual pseudopotential 
$-\beta w_{\rm b}(r)$, then $v_{\rm eff}(Q)$ in
Eq.~(\ref{eq:veffii}) becomes an effective ion-ion interaction in 
pseudopotential theory based on the linear-response formalism.

From the above formal and exact expressions 
(\ref{eq:QHNCii})-(\ref{eq:QHNCei}), the QHNC equations are obtained  by
introducing the following five approximations:\,\cite{QHNC}\\  
\begin{enumerate}
\item The electron-ion bridge-function is neglected in Eq.~(\ref{eq:QHNCei}): 
$B_{\rm eI}\simeq 0$ (the HNC approximation). 
\item The bridge-function $B_{\rm II}$ of the ion-electron mixture is
approximated by that of a one-component hard-sphere fluid (modified HNC 
-- MHNC -- approximation\cite{Ros79}).
\item The electron-electron DCF $C_{\rm ee}(Q)$ in the ion-electron mixture 
is approximated by that of the jellium model: $C_{\rm ee}(Q)\simeq -\beta v_{\rm ee}(Q)[1-G^{\rm jell}(Q)]$, which is written 
in terms of the the local-field correction (LFC) 
$G^{\rm jell}(Q)$. In our calculation we use the LFC proposed by 
Geldart and Vosko.\cite{GV}
\item An approximate $v_{\rm eI}(r)$ is obtained by treating 
a liquid metal as a nucleus-electron mixture\,\cite{NEmodel} in the form:
\begin{equation}
v_{\rm eI}(r)=-Z_{\rm A}\frac{e^2}{r}
 +\int v_{\rm ee}(|{\bf r}-{\bf r}'|)n_{\rm e}^{\rm b}(r')d{\bf r}'
 +\mu_{\mbox{\tiny XC}}[n_{\rm e}^{\rm b}(r)+n_0^e]
 -\mu_{\mbox{\tiny XC}}(n_0^{\rm e})\;.
\end{equation}
Here, $n_{\rm e}^{\rm b}(r)$ is the
bound-electron distribution and $\mu_{\mbox{\tiny XC}}(n)$ the
exchange-correlation potential in the local-density approximation, 
for which we have adopted the Gunnarsson-Lundqvist\cite{GL} formula.
Actually, the bound-electron distribution $n_{\rm e}^{\rm b}(r)$ 
is determined as the bound-electron part of $n_{\rm e}(r|U_{\rm e}^{\rm eff})\equiv n_{\rm e}^{0b}(r|U_{\rm e}^{\rm eff})+n_{\rm e}^{0f}(r|U_{\rm e}^{\rm eff})$, 
when the electron-ion RDF in Eq.~(\ref{eq:QHNCei}) is calculated from the wave equation for an electron under the external potential $U_{\rm e}^{\rm eff}(r)=
v_{\rm eI}(r)-{\it\Gamma}_{\rm eI}(r)/\beta$.  
\item The bare ion-ion interaction is taken as pure Coulombic: 
$v_{\rm II}(r)=Z_{\rm I}e^2/r$.  
\end{enumerate}
Under these approximations, a set of integral equations 
(\ref{eq:QHNCii})-(\ref{eq:QHNCei}) can be solved; 
its solution allows the determination of the electron-ion and ion-ion 
correlations together with the ionization and the electron bound-states.

For the MHNC closure relation we have used in this work the
parametrization for the bridge-function of a suitably chosen
hard-sphere reference system that was proposed recently by 
Rosenfeld.\cite{newbr} In this version of the MHNC the universality
hypothesis of the bridge-{\it function} \cite{Ros79} (which 'justifies'
the MHNC) is generalized to the level of the bridge-{\it functional} of
the reference system. This functional can be calculated very easily for
the general case of an inhomogeneous system of hard-spheres \cite{Ros89}
involving only fundamental measures; the functional is then specialized
-- as required for our case -- to a system of homogeneous hard-spheres.
This
fundamental measure bridge-functional is given in terms of characteristic
quantities of the {\it individual} spheres and
involves only integrations over known functions. Furthermore, in this
approach the
functional can be optimized by imposing the test-particle (or
source-particle) self-consistency, which is realized by the transition from an
inhomogeneous system to a homogeneous one if the source of the external
potential becomes a particle of the liquid.\cite{Per62} The
Ornstein-Zernike equation is then solved for the structure function
of the homogeneous system along with the closure relation where the
bridge-function is calculated by means of the above {\it functional},
assuming that the universality hypothesis is valid.
The obtained structure function is then fed into the bridge-functional 
yielding a new, improved bridge-function. This
procedure is iterated until numerical self-consistency is obtained in a sense 
that the structure function of the preceding step differs only
marginally from the present step. 

This method, which we denote by QHNC-MH, is in fact able 
to produce accurate data for the structure factors of Rb as demonstrated
in Fig.~\ref{fig:SQ313} in the direct comparison with 
experiments\cite{RbExp} at the triple point (313 K); the experimental
data are denoted by the open and black circles.

\section{RESULTS AND DISCUSSION}

In this section, we apply the QHNC-MH method to calculate the 
structure factors for compressed liquid-rubidium 
under high pressure (0.2, 2.5, 3.9, and 6.1 GPa), i.e., exactly the
same pressure-values under which 
Tsuji {\it et al.} \cite{tsuji} performed their experiments to measure
the structure factors. The corresponding densities 
are estimated by these authors to be 
1.07, 1.41, 1.56 and 1.95 times the normal density; 
the temperatures are 370, 520, 540 and 570 K, respectively.  

In a first step we apply our method to the 3.9 GPa state
and examine how accurately the MHNC equation (in combination with 
the Rosenfeld bridge-function) is able to reproduce the
RDF obtained in an MD simulation: 
the comparison is shown in Fig.~\ref{fig:RDFmd}, where the QHNC-MH 
result (full curve) is found to be undistinguishable from the MD RDF 
(open circles). This comparison confirms that the QHNC-MH method is able
to produce reliable structure-data that are as accurate as those
obtained in computer-experiments even for compressed liquids at high
densities; hence, MD simulations are no longer required in this study.

We can therefore proceed to compare the QHNC-MH results with {\it
experimental} data for all high-pressure states:
the structure factors calculated for pressures ranging from 0.2 to 6.1 
GPa are plotted by full curves in Fig.~\ref{fig:SQall} in comparison with the 
experimental results shown by full circles. We have found that 
our method gives structure factors in good agreement with experiment 
for 0.2 and 6.1 GPa.
However, in the case of 2.5 and 3.9 GPa, the agreement 
between the theoretical and experimental results is not so convincing:
a systematic 
deviation between the data sets is observed. In their analysis of their
experiment, Tsuji {\it et al.} \cite{tsuji} estimated 
the density of liquid Rb under pressure from 
the measured lattice constant of crystalline Rb including corrections for 
the thermal expansion and the volume jump at melting.
However, according to Tsuji,\cite{tsujiprv} there remains 
an uncertainty in the evaluation 
of the density, that might be responsible for these deviations.
Therefore, we have modified the density for the 3.9 GPa case: 
when decreasing the ion-sphere radius (Wigner-Seitz radius)
$a$ [$a = \left(3 V/4 \pi \right)^{1/3}$ and $V$ being the specific volume per
ion] by a factor of 1/1.05 
we can find indeed a good agreement between the theoretical 
and experimental results for the 3.9 GPa state, as shown in 
Fig.~\ref{fig:SQrees}; 
a similar good agreement is found also for the 2.5 GPa case 
if the Wigner-Seitz radius $a$ is decreased by 5 \%.

It should be mentioned that the concept of 
our QHNC-method is for high pressures  
as reliable and valid
as for the room pressure, since all the approximations 
entering this method remain valid as the pressure is increased. 
This fact can be seen from the result for the electronic structure, which 
will be discussed in the following.
The bound levels of the ion in compressed liquid Rb are almost 
the same as 
those at room pressure. As a consequence, the electron-ion RDF's remain 
 -- as displayed in Fig.~\ref{fig:RDFei}  -- almost 
unchanged for the five states considered.
The electron-ion DCF $C_{\rm eI}(r)$ is determined by Eq.~(\ref{eq:QHNCei}). 
Figure~\ref{fig:Cei} illustrates the pressure-variation of the 
electron-ion DCF $C_{\rm eI}(r)$, which -- as noted above --
plays the role of a non-linear 
pseudopotential in the expression for the effective ion-ion interaction
[cf. Eq. (2.5)]; also the electron-ion DCF does not change significantly 
under these pressure-variations. 

Here, note that the electron-ion structure factor $S_{\rm eI}(Q)$, 
the Fourier transform of the electron-ion RDF, is written in the following form:
\begin{eqnarray}
S_{eI}(Q)&=&\sqrt{n_0^In_0^e}\,C_{eI}(Q)\chi_Q^0 / D(Q) \label{e:sei1}\\
         &=& {\rho(Q) \over \sqrt{Z_{\rm I}} }S_{II}(Q)\, ,   \label{e:sei2}
\end{eqnarray}
where
\begin{eqnarray}
\rho(Q)&\equiv& {n_0^e C_{eI}(Q)\chi_Q^0 \over 1-n_0^e C_{ee}(Q)\chi_Q^0 }\;
 \label{e:qrho}\;,\\
D(Q)&\equiv& [1-n_0^I C_{II}(Q)][ 1-n_0^e C_{ee}(Q)\chi_Q^0 ]-n_0^In_0^e| 
C_{eI}(Q)|^2\chi_Q^0 . 
\end{eqnarray}
Hence, Eq.~(\ref{e:sei2}) can be represented in $r$-space as
\begin{equation}
n_0^e g_{eI}(r) = \rho(r) +n_0^I\int \rho(|{\bf r}-{\bf r'}|)\,g_{II}(r'){\bf r'}, \label{e:sps}
\end{equation}
which states that the free-electron distribution $n_0^eg_{eI}(r)$ around 
an ion can be described exactly by the superposition of surrounding 
``neutral pseudoatoms". Each ion carries a screening electron-cloud 
$\rho(r)$ [with $\int \rho(r)d{\bf r}=Z_{\rm I}$], and makes it thus 
electrically neutral (including the core-electrons) as if it were an atom. 
Therefore, in this formalism a liquid metal can be considered to be composed 
by neutral pseudoatoms.
Using Eq.~(\ref{e:qrho}), the free-electron density distribution 
of a pseudoatom is calculated for liquid Rb under high pressures 
(cf. Fig.~\ref{fig:pseudatom}): the results indicate that 
the electron-density distribution of 
a pseudoatom suffers no significant change outside of the core-region (where the bound-electron density is large) 
under this pressure-variation. 
Summarizing, we can conclude that a pseudoatom remains almost unchanged
in comparison to room pressure even when the density is increased by a
factor of nearly 2 (high pressure-state 6.1 GPa). 
On the other hand, Tsuji {\it et al}. expected that a hard-sphere model\cite{AL} might be 
successfully applied to describe the structure of liquid alkali metals and that the effective diameter of the hard-sphere should vary with pressure due to the change in the screening effect; that is, the size of a pseudoatom is assumed to be changed with pressure.
Based on this model, the effective hard-sphere radius $\sigma_{\rm H}$ is considered to vary under pressure-variation keeping the packing fraction 
$\eta=\pi n_0^{\rm I}\sigma_{\rm H}^3/6$ to be constant at a value of 
0.45; this leads to a uniform-compression model in which the 
position of first peak $Q_1$ in the structure factor should 
increase proportionally to $({n_0^{\rm I}})^{1/3}$  
with increasing pressure.

We also find in our approach that 
the effective ion-ion interaction for liquid Rb is 
practically invariant under this pressure-variations (cf. Fig. 8),
i.e., it remains almost the same as the one at room pressure. 
In contrast to the hard-sphere model, where the essential repulsive part
of the effective ion-ion potential should be scaled in units of $a$, 
no scaling feature is observed for the effective ion-ion interaction 
in units of the Wigner-Seitz radius $a$.
Nevertheless, 
all the structure factors for all these five pressure-values almost
coincide in one single curve when scaling the wavenumber 
in units of the Wigner-Seitz radius $a$ (as shown in Fig.~\ref{fig:scaledSQ}).

Even if the liquid-rubidium potential under these pressure-variations 
can be considered as invariant with respect to the one at 
room  pressure (neglecting small deviation), we have also observed the above mentioned scaling property 
in the structure factor (cf. Fig.~\ref{fig:scaledSQ}); 
this means that the liquid-rubidium potential has a very special feature.
For a {\it general} liquid (the Lennard-Jones 
potential for example), it is impossible to display the structure
factor of two different states, scaled in length-units 
in such a way that they
practically coincide. For a liquid alkali metal such a scaling property
{\it can} be observed: to examine this, 
let us consider liquid Li as a further example.
At first we calculate the effective ion-ion interaction of liquid Li at room 
pressure (470 K, $n_0$ with $r_{\rm s}=3.308$), and assume -- as
for Rb -- that the effective Li potential 
is unchanged with respect to the one at room pressure.
Then, we can easily find 
the following three sets of temperatures and densities: (470 K, $n_0$), 
(600 K, 1.34$n_0$) and (750 K, 2.37$n_0$), i.e. states for 
which the structure 
factors can be scaled almost in a single curve in unit of $a$, 
as demonstrated in Fig~\ref{fig:LiSQ}. For alkali liquid metals, 
this scaled structure 
factor is not very sensitive to states specified by the plasma parameter $\Gamma\equiv \beta e^2/a$ and $r_s\equiv a/a_{\rm B}$ 
in the sufficiently high-density region. 
This is the main reason 
why structure factors of all alkali liquids (from Li to Cs) 
near the triple point coincide almost in a single curve 
when scaling $Q$ in units of the Wigner-Seitz radius $a$, 
as was observed experimentally\cite{ExpNaKCs,Waseda} and 
theoretically.\cite{Minoo,fpMD,Baluc} 

This scaling property in the structure factors of liquid Rb signifies 
that the first peak of the structure factor appears almost 
at the same position $Q_1a$ (in scaled units) for all these pressures; 
this means that 
the position of the first peak $Q_1$ in the structure factor (taken in absolute values) should 
increase proportionally to $x=(V/V_0)^{-1/3}$, where $V_0$ is the specific volume at
room pressure. 
In our calculation, this peak-position in reduced units is estimated from Fig.~\ref{fig:scaledSQ} to be $Q_1a=4.30$, 
from which we obtain the relation $Q_1=1.51x$ because of 
$a=r_{\rm s}a_{\rm B}=2.58/x$; this linear relation is plotted in 
Fig.~\ref{fig:Q1unif}.
In this figure, the full and the open circles denote the experimental points 
obtained for several states by Tsuji {\it et al.};\cite{tsujiprv2} in
particular, the five full circles represent those 
states which we have investigated in our theoretical study, i.e., 
for pressure-values 
0, 0.2, 2.5, 3.9, and 6.1 GPa, respectively. 
These {\it experimental} $Q_1$-points in Fig.~\ref{eq:QHNCei} are close
to the linear relation which was determined from the {\it calculated} 
structure factors; thus, this figure demonstrates that 
the uniform-compression model (corresponding to a 
linear relation denoted by the full line) is
indeed valid, although there is some uncertainty 
due to difficulties in estimating the density and to problems in experiments 
under high pressure, in general. 

\section{CONCLUSIVE DISCUSSION}
The QHNC method in combination 
with the Rosenfeld bridge-functional has been shown 
to reproduce the experimental structure factors of liquid Rb under 
high pressures (ranging from 0 to 6.1 GPa) very accurately; for the case
of 2.5 and 3.6 GPa the experimentally estimated densities have to be
readjusted in terms of a 5 \% variation of the Wigner Seitz radius
$a$. 
Furthermore we observe that the structure factors coincide almost in one single curve if wavenumber are scaled in units of
$a$ (Fig.~\ref{fig:scaledSQ}). 
This indicates clearly that liquid Rb is uniformly compressed as 
the pressure is increased: this, in turn signifies, that
the first peak-position $Q_1$ of the structure 
factors increases proportional to $(V/V_0)^{-1/3}$ (Fig.~\ref{fig:Q1unif}). 

It must be mentioned that in contrast to our result, Shimojo 
{\it et al.}\cite{Shimojo} conclude from their result obtained in 
the first-principles MD simulations that some deviation from uniform
compression exists for the 6.1 GPa state, though liquid 
Rb is compressed uniformly at 2.5 GPa.
Their experimental RDF's for 2.5 and 6.1 GPa are obtained by a Fourier transform 
of the experimental structure factors of Tsuji {\it et al}.
On the basis of the RDF's they observe a different behaviour than
on the basis of the structure factor, 
although they have used the same experimental structure factors and 
Rb states as we did: (i) the first-peak position in the 
RDF at 2.5 GPa follows the uniform compression model, while the first-peak 
position in the RDF at 6.1 GPa shows a deviation from the uniform 
compression model. Thus, they have asserted that their calculated 
result agrees with the experiment.
(ii)
In contrast 
to their RDF's, the experimental structure factor at 2.5 GPa shows 
a substantial deviation from the uniform compression
compared to the 6.1 GPa state; the position $Q_1$ of the first peak in 
$S_{\rm II}(Q)$ lies far away from the uniform compression line in 
Fig.~\ref{fig:Q1unif} while on
the other hand, the experimental structure factor at 6.1 GPa 
shows that the uniform compression model is still valid: the 
$Q_1$-point for 6.1 GP in Fig.~\ref{fig:Q1unif} is very close to the
uniform compression line.

According to Tsuji,\cite{tsujiprv} this discrepancy between the conclusions
based on the RDF's and the structure factors 
is possible, since there is some experimental ambiguity in the value 
of the peak height in the structure factor while the peak-position is 
accurate and reliable. Nevertheless, 
from all the experimental data displayed in Fig.~\ref{fig:Q1unif}, we can 
conclude that liquid Rb is compressed uniformly up to a pressure of 6.1 GPa; 
however, a more detailed discussion if deviations
from the uniform compression model (cf. Fig.~\ref{fig:Q1unif}) have any 
physical meaning is not very conclusive, due to the uncertainty in the 
evaluation of the experimental density and the difficulties encountered
in experiments under high pressure.

It is interesting to notice that the structure factors of compressed 
liquid Rb coincide practically in one single curve if wavenumbers are scaled 
in units of the Wigner-Seitz 
radius $a$, despite of the fact that the effective ion-ion interaction
remains under pressure 
unchanged with respect to room pressure; this means that interaction potentials of liquid alkali metals have a special characteristic property, 
as we have demonstrated and confirmed in addition
for the case of liquid Li in 
Fig.~\ref{fig:LiSQ}. The neutral pseudoatom in compressed liquid 
Rb remains almost unchanged under pressure-variation, which, in
turn, is the reason why the effective ion-ion interaction remains practically 
invariant, similar to state-{\it independent} interactions, such as 
the one for liquid argon.

We have demonstrated 
in this contribution that the QHNC method can treat the 
'outer-structure' problem (i.e., calculation of
the ion-ion and electron-ion RDF's) and 
the 'inner-structure' problem (i.e., calculation of the
electronic structure of the ion) in a 
self-consistent way using the atomic number as the only input data: 
therefore, this method can be considered to be very 
useful to treat a plasma, where the ionization 
$Z_{\rm I}$ is not known beforehand and where there is no way of
constructing a 
pseudopotential to give the
effective ion-ion interaction.
From the successful results for compressed liquid Rb presented
here we can conclude that the QHNC method 
is expected to be nicely
applicable for plasma states in a wide range of densities and 
temperatures.

\acknowledgments

We thank Professor Tsuji and Dr. Morimoto for sending us numerical experimental data 
and for discussions on the analysis of the experimental data. GK
acknowledges financial support by the \"Osterreichische Forschungsfonds
under Project No. P11194-PHY.


%
%

\begin{figure}
\caption{Ion-ion structure factor $S_{\rm II}(Q)$ for liquid Rb 
at a temperature of 313 K; the QHNC-MH method (in combination with the 
the Rosenfeld bridge-functional) yields a structure factor (full curve) 
in excellent agreement with experiments\cite{RbExp} 
(open and full circles).}
\label{fig:SQ313}
\end{figure}

\begin{figure}
\caption{Ion-ion RDF $g_{\rm II}(r)$ at a pressure of 3.9 GPa (full curve) 
calculated by the QHNC-MH method: results are undistinguishable  
from those obtained in MD simulations (open circles).}
\label{fig:RDFmd}
\end{figure}

\begin{figure}
\caption{Structure factors $S_{\rm II}(Q)$ of liquid Rb under high pressures: 
0.2, 2.5, 3.9 and 6.1 GPa; the QHNC-MH results (full curve) are 
compared with experiment\cite{tsuji} (full circles). The densities 
corresponding  
to these pressures are 1.07, 1.41, 1.56 and 1.95 
times the normal density $n_0$, respectively.}
\label{fig:SQall}
\end{figure}

\begin{figure}
\caption{The reestimated structure factor $S_{\rm II}(Q)$
(full curve) of liquid Rb at 3.9 GPa 
where the Wigner-Seitz radius $a$ has been decreased by a factor of 
1/1.05 as a unit of length; excellent agreement with experimental
results\cite{tsuji} (open circles) is observed}
\label{fig:SQrees}
\end{figure}

\begin{figure}
\caption{The electron-ion RDF's $g_{\rm eI}(r)$ of liquid Rb under pressure 
(ranging from 0 to 6.1 GPa); note that the $g_{\rm eI}$'s 
remain almost unchanged under pressure-variation.}
\label{fig:RDFei}
\end{figure}

\begin{figure}
\caption{The electron-ion DCF's $C_{\rm eI}(r)$ of liquid Rb under pressure
(ranging from 0 to 6.1 GPa); these functions play the role of a nonlinear 
pseudopotential $w_{\rm b}(r)$ 
to determine the effective ion-ion interaction.}
\label{fig:Cei}
\end{figure}

\begin{figure}
\caption{The electron-density distribution $\rho(r)$ 
of a pseudoatom in liquid Rb 
under pressure (ranging from 0 to 6.1 GPa); the 5s-electron density 
distribution of a free atom is displayed for comparison (small full
circles). $\rho(r)$ is plotted in units of $a_{\rm B}^{-3}$ ($a_{\rm B}$
being the Bohr radius).}
\label{fig:pseudatom}
\end{figure}

\begin{figure}
\caption{The effective ion-ion interaction $v_{\rm eff}(r)$ of liquid Rb 
under pressure (ranging from 0 to 6.1 GPa); the effective interaction for 
compressed liquid Rb remains practically unchanged with respect to 
room pressure.}
\label{fig:effVii}
\end{figure}

\begin{figure}
\caption{Structure factors $S_{\rm II}(Q)$ of liquid Rb calculated for 0, 0.2, 
2.5, 3.9, and 6.1 GPa; all results are scaled in units of the Wigner-Seitz
radius $a$ and practically coincide in one curve; the experimental 
result\cite{tsuji} 
for 6.1 GPa is plotted by full circles.}
\label{fig:scaledSQ}
\end{figure}

\begin{figure}
\caption{Structure factors $S_{\rm II}(Q)$ of liquid Li calculated for 
two states, ($1.34n_0$, 600 K) and ($2.34n_0$, 750 K) under 
the assumption that the effective ion-ion interaction is the same
as that for room pressure. The structure factor 
(full curve) of liquid Li at room pressure is also plotted 
for comparison; when scaled in units of the Wigner-Seitz radius $a$, 
all three structure factors practically coincide in a single curve.}
\label{fig:LiSQ}
\end{figure}

\begin{figure}
\caption{The position of the first peak $Q_1$ in the structure factor for 
liquid Rb as a function of $(V/V_0)^{-1/3}$; open and full circles 
represent experimental results.\cite{tsujiprv2} 
In particular, the full circles are the points 
determined from the experimental structure factors under pressure: 
0, 0.2, 2.5, 3.9, and 6.1 GPa plotted in Figs.~\ref{fig:SQ313} and \ref{fig:SQall}. 
The solid line denotes our calculated results (uniform compression
model) derived from the first peak-position ($Q_1a=4.30$) 
in the scaled structure factor in Fig.~\ref{fig:scaledSQ}.}
\label{fig:Q1unif}
\end{figure}

%
%


\begin{references}
\bibitem{tsujiprv2} The experimental data for the $x-Q_1$ relation are
provided by Tsuji in the numerical table.
\end{references}
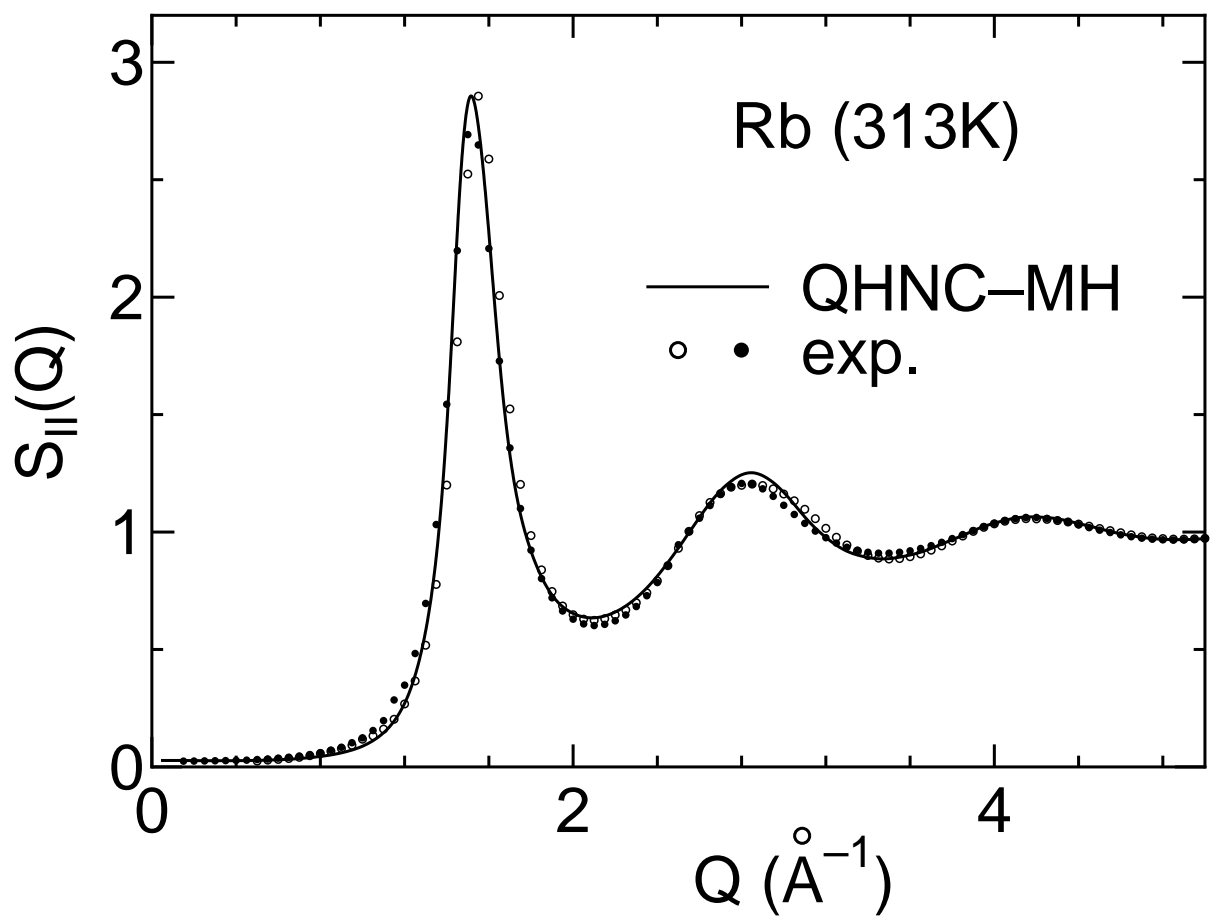
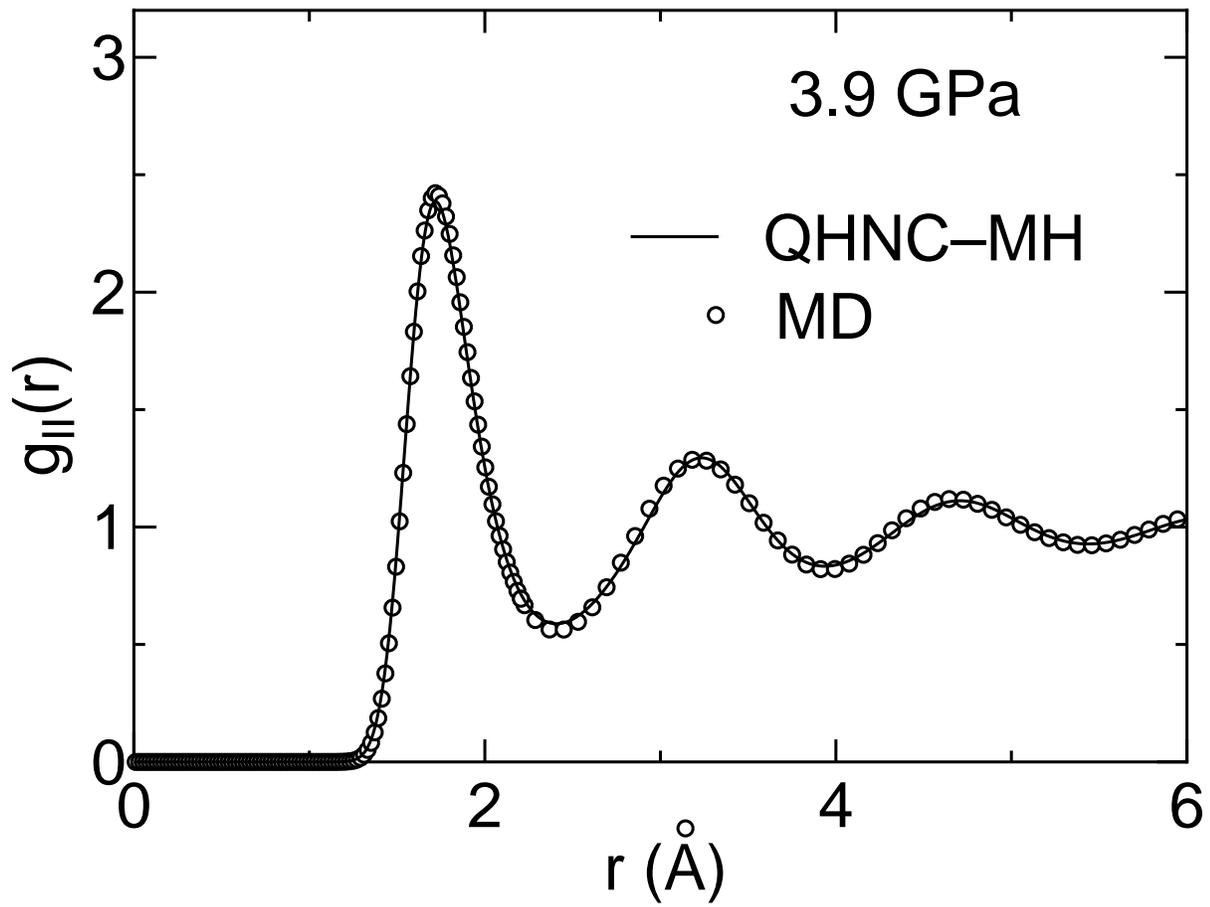
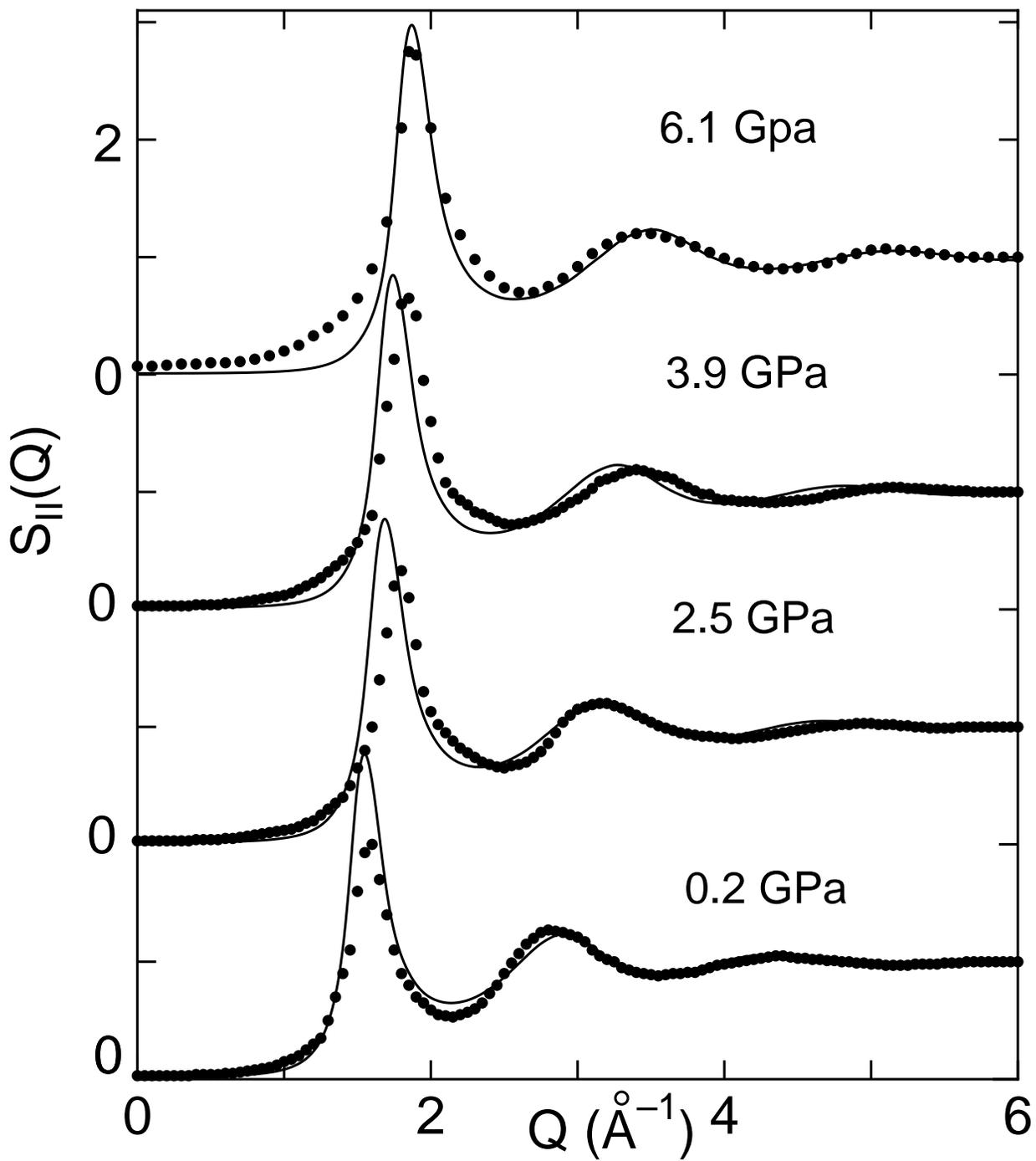
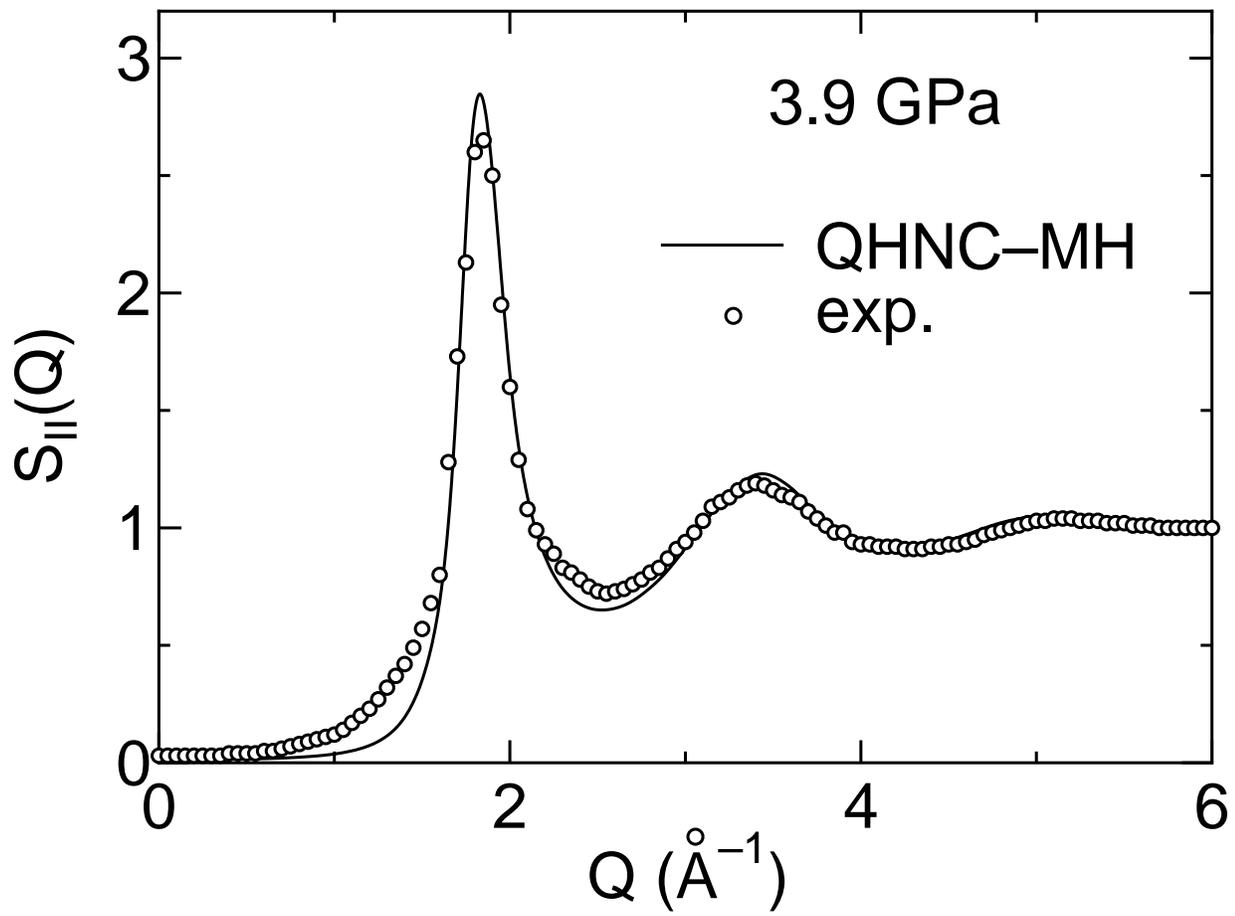
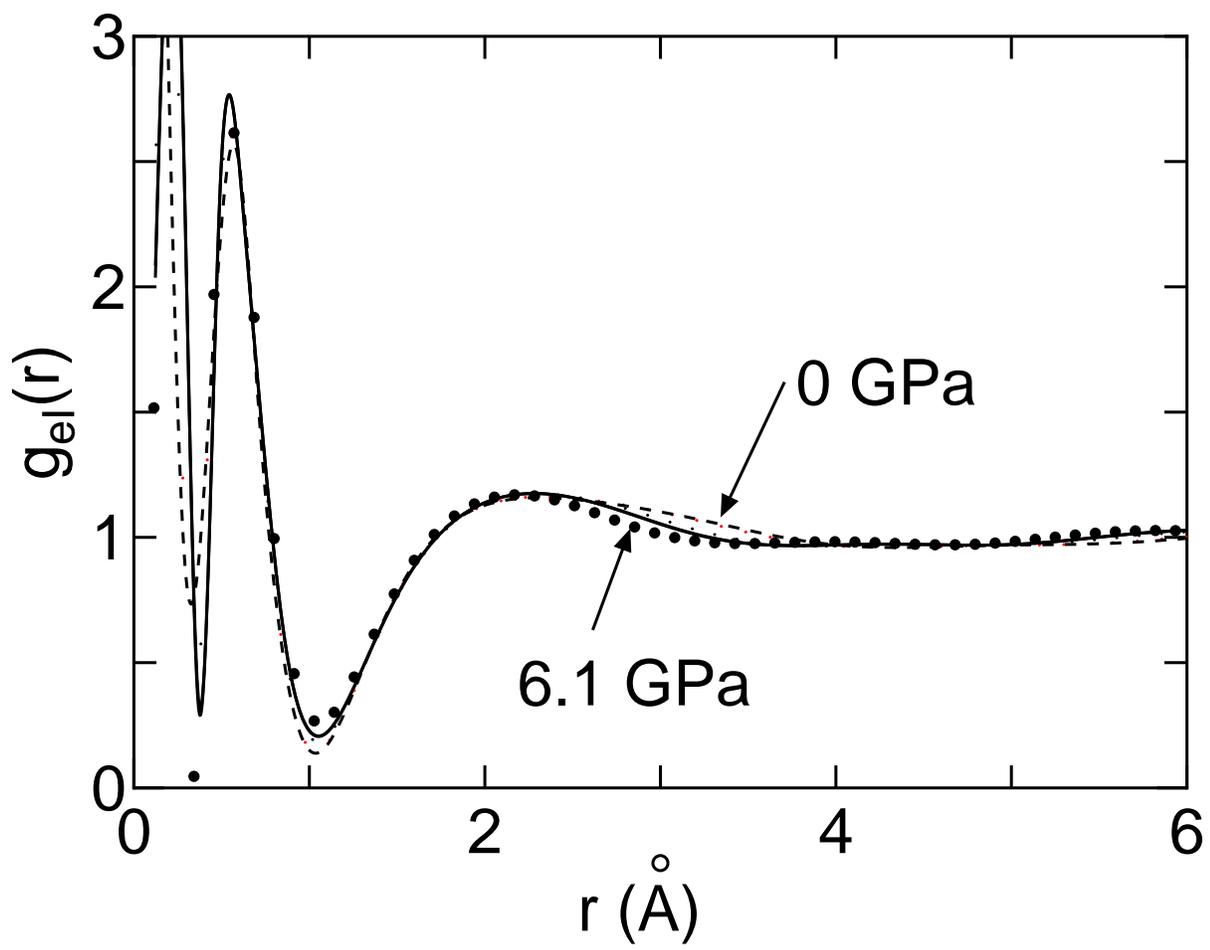
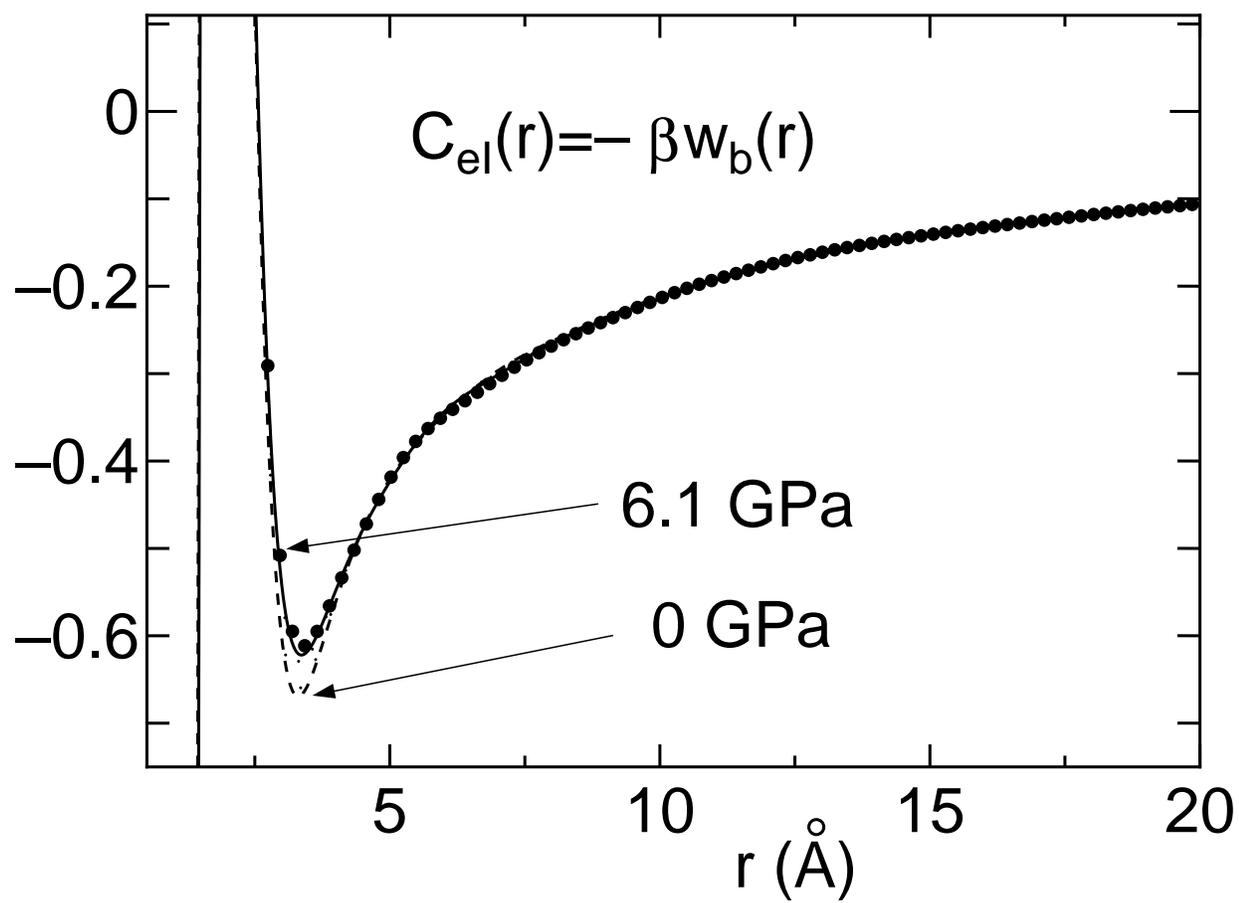
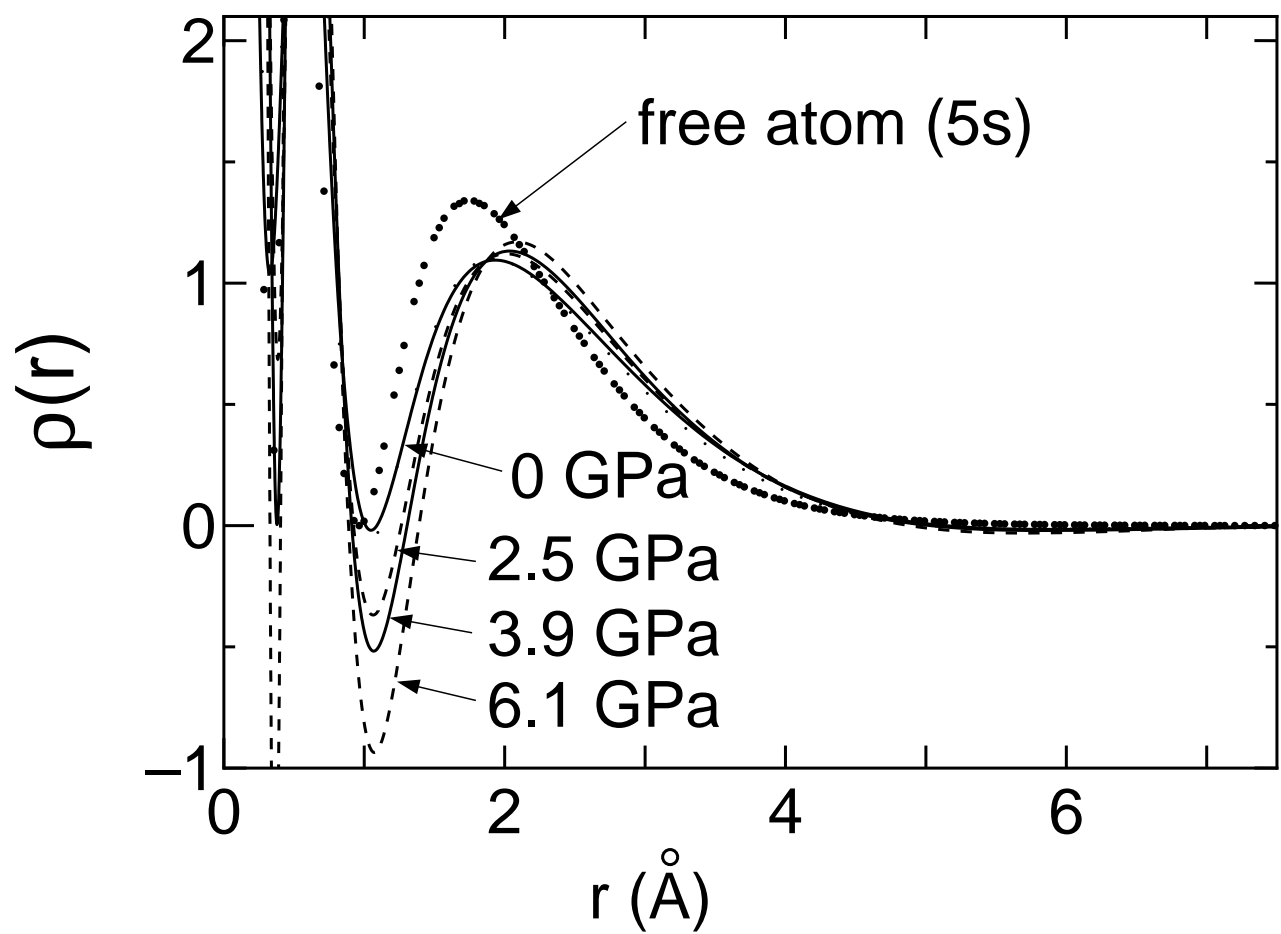
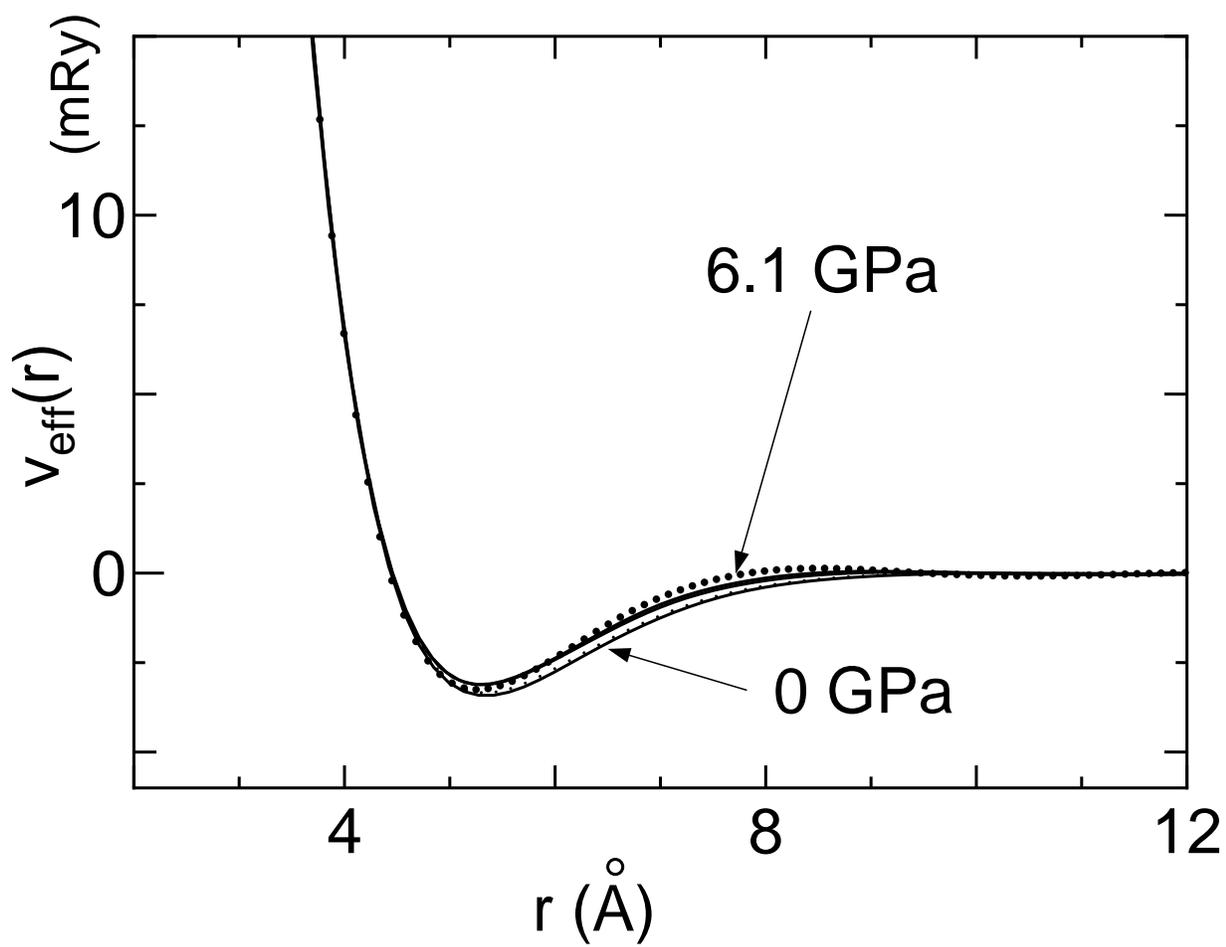
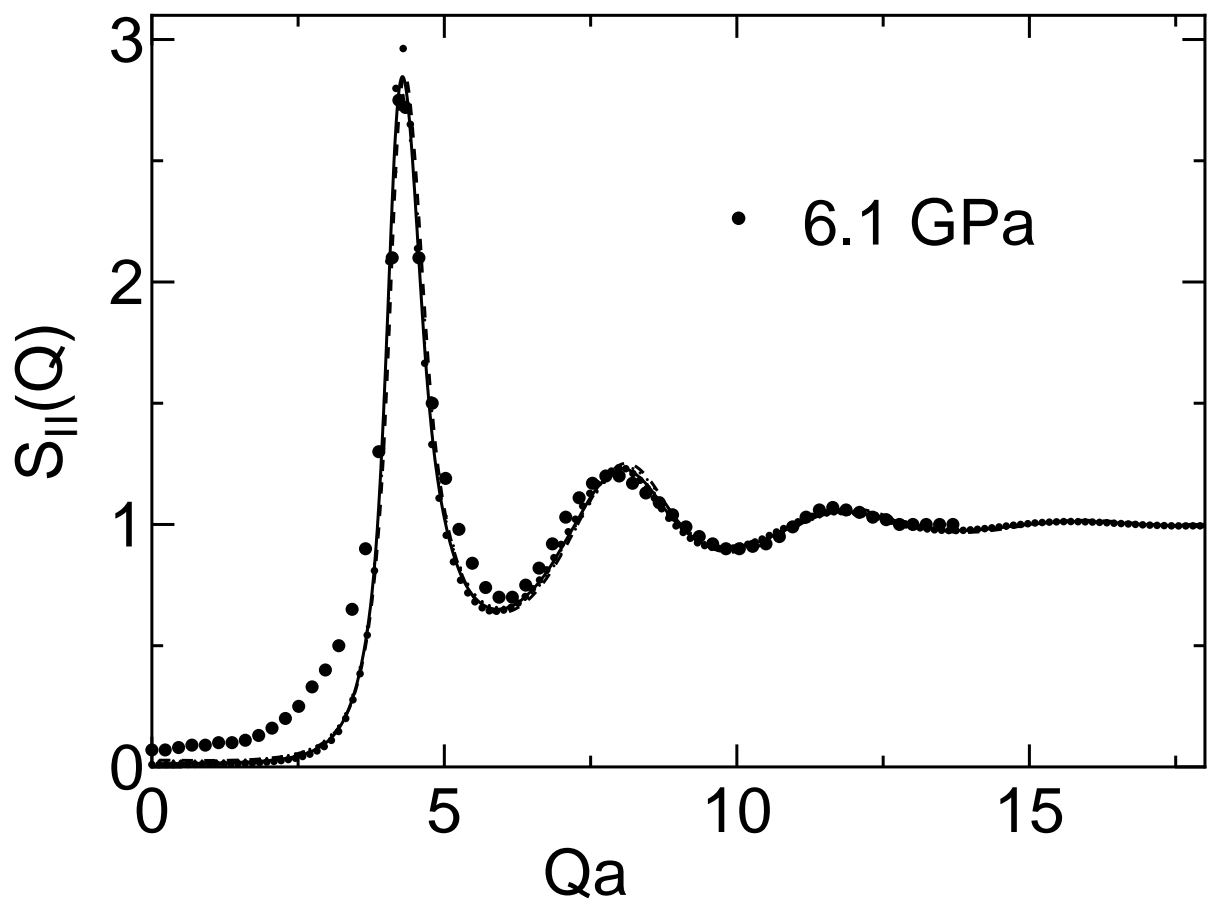
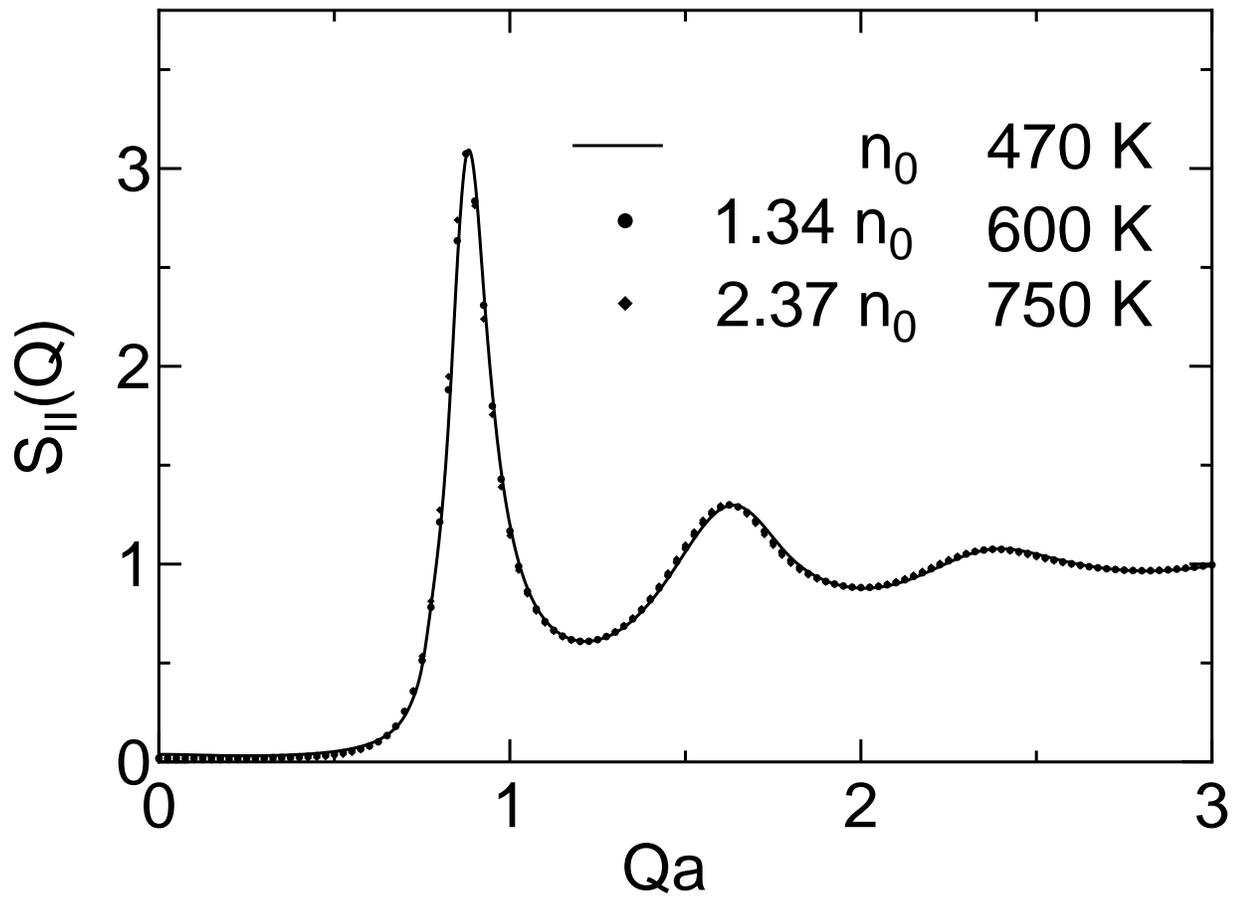
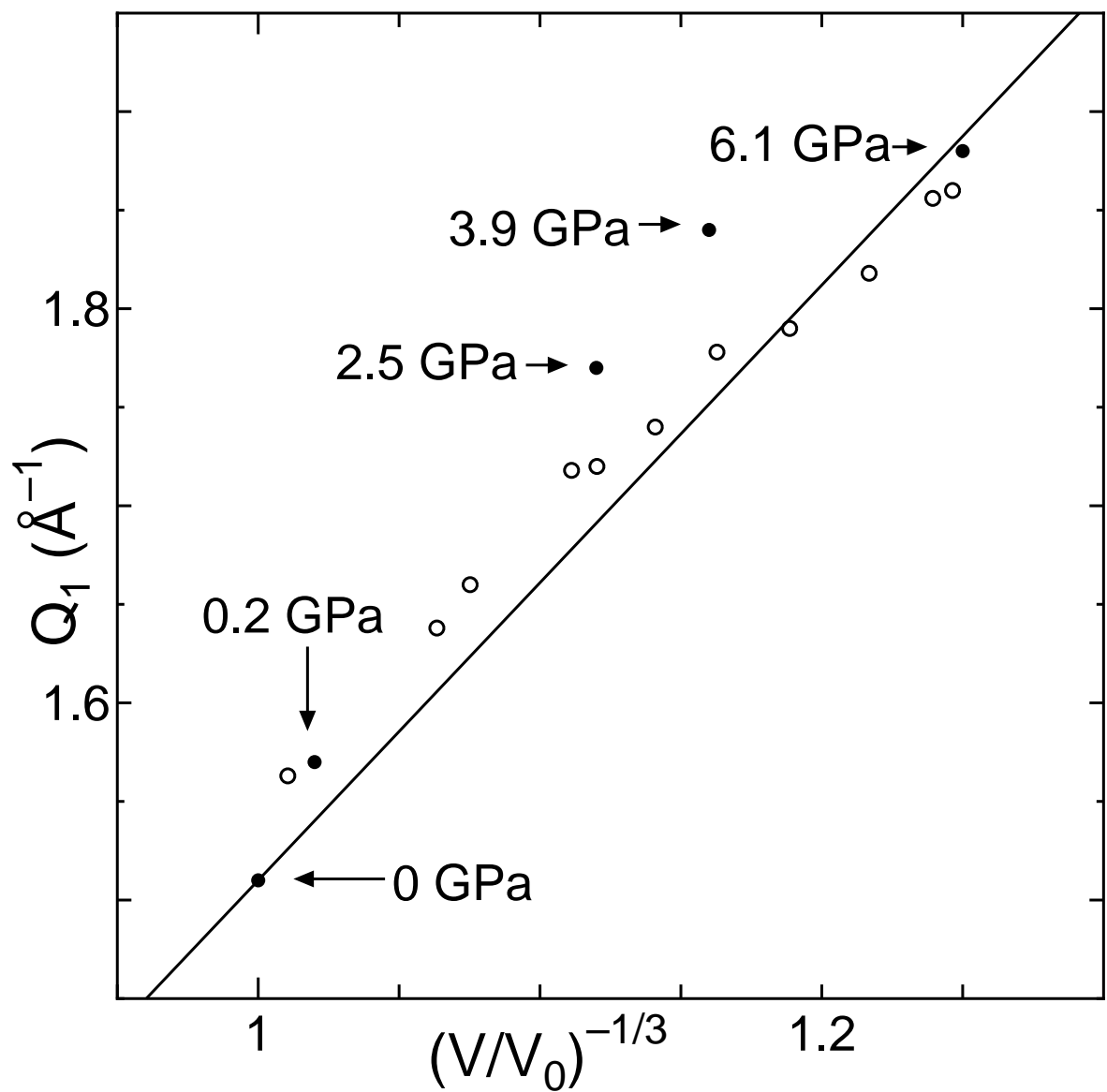
\end{document}